\documentstyle[12pt,epsf]{article} 
\newcommand{\ba}{\begin{array}}
\newcommand{\ea}{\end{array}}
\newcommand{\bd}{\begin{displaymath}}
\newcommand{\ed}{\end{displaymath}}
\newcommand{\be}{\begin{equation}}
\newcommand{\ee}{\end{equation}}
\newcommand{\bea}{\begin{eqnarray}}
\newcommand{\eea}{\end{eqnarray}}

\def\bra{\langle}
\def\ket{\rangle}

\def\a{\alpha}
\def\b{\beta}

\def\e{\epsilon}

\def\l{\lambda}

\def\n{\nu}

\def\pr{\prime}

\begin{document}

\begin{flushright}
\tt{hep-ph/0107279} 
\end{flushright}
\vskip 5pt
\begin{center}
{\Large {\bf Anomalous magnetic moment of muon and L-violating
Supersymmetric Models}}
\vskip 20pt

\renewcommand{\thefootnote}{\fnsymbol{footnote}}

{\sf Rathin Adhikari $^{a,\; b}$
\footnote{E-mail address: rathin\_adhikari@yahoo.com}},   
and
{\sf G. Rajasekaran $^{c}$
\footnote{E-mail address: graj@imsc.ernet.in}

\vskip 10pt  
$^a${\it Physics Department, University of Calcutta,
Kolkata (Calcutta) 700009, India}\\ 

$^b${\it Physics Department, Jadavpur University, Kolkata (Calcutta)
 700032, India}\\

$^c${\it The Institute of Mathematical Sciences, 
Chennai (Madras) 600113, India}}

\vskip 15pt
{\bf Abstract}
\end{center}

{\small 
We consider L-violating Supersymmetric Models  to explain the
recent muon $g_{\mu} -2$ deviation from the Standard Model. The
order of trilinear L-violating couplings which we require 
also generate neutrino mass  which is somewhat higher than
expected unless one considers highly
suppressed $L-R$ mixing of
sfermions. However, without such fine tuning for sfermions it is 
possible to
get appropriate  muon $g_{\mu}-2$ deviation as well as neutrino
mass if one considers some horizontal symmetry for the lepton
doublet. 
Our studies show that  $g_{\mu} -2$  deviation may not imply
upper bound of about 500 GeV on masses of supersymmetric
particles like chargino
or neutralino as proposed by other authors for $R$ parity
conserving supersymmetric models. However, in our scenario
sneutrino mass is expected to be light ($\sim 100$ GeV) and 
$e-\mu-\tau$ universality violation may be observed
experimentally in near future.
}

\vskip 20pt
\begin{center}
PACS NO. 12.60.Jv, 13.40.Em, 14.80.Ly 
\end{center}

\newpage

\setcounter{equation}{0}
There is recent indication \cite{brown} that the anomalous
magnetic moment of muon (AMMM) differs from its Standard Model value by 
$a_{\mu}^{exp}-a_{\mu}^{SM} = 42 (16) \times 10^{-10}$. If it is
further confirmed in future experiments this will be a
strong evidence of new physics beyond the Standard Model apart
from the other evidence coming from the neutrino oscillation
experiments \cite{bach,solar1,solar2,atm} indicating massive neutrinos. 
Both these evidences of new physics - anomalous magnetic moment of muon and
neutrino mass , might be related in some models like in the leptonic 
Higgs doublet model
considered in \cite{ma} or in $R$-parity
violating Supersymmetric Models which we are considering here. 

There are several explainations of this recent AMMM experiment
\cite{brown} in the context of Supersymmetric
Models \cite{msusy,kim} and also in the context of
non-Supersymmetric Models \cite{nonsusy}. 
Recently, $R$ parity violation in Effective Supersymmetry has been
considered \cite{kim} to explain the deviation in $a_{\mu}$.  To
avoid the constraint coming from muon neutrino mass they have
considered particularly the semileptonic $L$ violating couplings
$\l^{\pr}_{213}$ and $\l^{\pr}_{231}$ instead of $\l^{\pr}_{233}$.
However the product of these couplings also contribute to muon 
neutrino mass. Unless one considers highly
suppressed $L-R$ mixing of sfermions in general it is difficult
to satisfy the general upper   
bound \footnote{This follows 
if one consideres the
bound from Tritium beta decay as well as the mass squarred
differences of different flavor of active neutrinos taking part
in oscillation \cite{bach}.} on active neutrino mass $\sim 2.2$ eV.

In this letter we have shown that 
in $R$ parity violating Supersymmetric model considering upper 
bound on the neutrino masses and also
considering the recent observation of $\Delta a_{\mu}$
\cite{brown} one gets constraint on soft
susy breaking parameters due to the requirement of suppression of
$L-R$ mixing of sfermions. However, this indicates certain amount
of fine tuning in the SUSY parameters. On the other hand, 
it seems natural to consider some
horizontal symmetry \cite{vol,marc,babu,bar} 
for the lepton doublet superfields in the $R$
violating Supersymmetric Model for which it is
possible to get
simultaneously large $\Delta a_{\mu}$ as observed
\cite{brown} as well as
the light neutrino mass satisfying the present upper bound of
about 2.2 eV.
Furthermore, in this scenario only sneutrino mass is required to
be light about 100 GeV whereas unlike $R$ conserving
Supersymmetric Model there are no 
upper bounds on masses of 
other supersymmetric particles like chargino or neutralino 
\cite{msusy}.

We shall discuss first how in $R$ violating Supersymmetric
Model  \cite{rvsusy}  anomalous magnetic moment of muon get some
contribution through $L$ violating couplings and how in such
scenario majorana neutrino mass is also generated.
In $R$ parity  violating Supersymmetric Model as the three lepton
supermultiplet $L_m \;\; m=1,2,3$ and down-type Higg 
supermultiplet transform
identically we denote these four supermultiplets as $L_{\a} \;\;
\a=0,1,2,3$. Imposing $Z_3$ baryon triality one can write the most
general 
renormalizable $L$ violating terms in the
superpotential \cite{rvsusy} as:
\bea
W=\e_{ij}  \left[ -\mu_{\a} L_{\a}^i H_U^j + {1 \over 2} \l_{\a
\b m} L_{\a}^i L_{\b}^j E_m + \l_{\a n m }^{\pr} L_{\a}^i Q_n^j
D_m - h_{nm} H_U^i Q_n^j U_m \right]
\eea
in which the coefficients $\l_{\a \b m}$ are antisymmetric under
the interchange of the indices $\a$ and $\b$. 
$H_U$ is the up-type Higgs supermultiplet, $Q$ are chiral superfields 
containing quark 
doublets and $U_m$ and $D_m$ are the singlet up and down type
quark supermultiplets respectively and $E_m$ are the singlet
charged lepton supermultiplet.   Some soft supersymmetry breaking
terms relevant for our discussion are
\bea
V_{soft} &=& {\left( M_{\tilde L}^2 \right)}_{\a \b} \tilde{{L_{\a}^i}^*}
\tilde{L_{\b}^i} + {\left( M_{\tilde E}^2 \right)}_{m n} \tilde{{E_{m}}^*}
\tilde{E_{n}}  - \left( \e_{ij} b_{\a} \tilde{L_{\a}^i} H_U^j + H.c
\right) + \e_{ij}  \left(  {1 \over 2} a_{\a \b m}
\tilde{L_{\a}^i}   \tilde{L_{\b}^j}  \tilde{E_m} \right. \nonumber \\ 
&+& \left. a^{\pr}_{\a n
m} \tilde{L_{\a}^i} \tilde{Q^j_n} \tilde{D_m} - (a_U)_{nm} H_U^i
\tilde{Q^j_n} \tilde{U_m} + h.c. \right)
\eea
in which the single $B$ term of the $R$ conserving minimal
Supersymmetric standard Model (MSSM) has been extended to the 4
component vector $b_{\a}$ . Similarly $A$ parameters of MSSM has
been extended to $a_{\a \b m}$ and $a^{\pr}_{\a n m}$. Here $a_{\a \b
m}$ are antisymmetric in the first two indices $\a$ and $\b$.  
In a convenient notation  such $A$ and $B$ parameters have been
written above as
\bea
a_{\a \b m} = \l_{\a \b m} (A_E)_{\a \b m} \; \; , a^{\pr}_{\a n m} =
\l_{\a n m} (A_D)_{\a n m} \; ,  
\nonumber
\eea
\bea
(a_U)_{nm} = h_{n m} (A_U)_{n m} \;\; , b_{\a} =
\mu_{\a} B_{\a} 
\nonumber
\eea
We assume that only the neutral scalar fields acquire vacuum
expectation values and we write $\bra h_U  \ket = v_u/\sqrt{2}$
and $ \bra \tilde{ \nu_{\a}}\ket = v_{\a}$.  If 
$\mu_{\a}$ and $v_{\a}$ are almost aligned at
the low energy scale  by some mechanism  then only the $\mu_i$
(where $i = 1..3$) can be rotated away. However, if such
alignment condition is achieved at some scale of supersymmetry
breaking normally it breaks down at the low energy scale. Due to
this misalignment 
parametrised as $\xi_i = v_i/v_0 - \mu_i/\mu_0$,
particularly the mixing of neutralino/neutrino, chargino/charged
lepton and slepton/Higgs mixing occur \cite{hemp}.  We
shall neglect such mixing considering $\mu_i$ to be very small
and also we neglect the contribution to AMMM coming from
the effective trilinear terms obtained from bilinear terms with
suitable redefinition of $L$ and $H$ superfields.
To get appreciable $\Delta a_{\mu}$ we need large
trilinear $\l_{ijk}$ or $\l^{\pr}_{ijk}$ couplings as will be
shown later.

As contribution to AMMM comes also due to mixing between
$L$ type and $R$ type    charged slepton/d-squarks in the Feynman
diagrams we discuss
this mixing in brief for any particular generation. 
The $L-R$ unitary mixing 
matrix is given by
\bea
U^{\tilde l} &=&\left(
\begin{array}{cc} \cos \phi &  \sin \phi \\
-\sin \phi & \cos \phi \\
\end{array}
\right)
\eea
corresponding to the $2 \times 2 $  $LR$ block
of the charged slepton  squarred mass matrix 
\bea
M^2({\tilde l}) &=&\left(
\begin{array}{cc} L^2 + m_l^2 &  A m_l \\
A m_l & R^2 + m_l^2 \\
\end{array}
\right)
\eea
where $L^2 =   {M^2(1)}_{ll}  + (T_3 - e \;\sin^2 \theta_w)\;
m_z^2 \;\cos 2 \beta$, $ R^2 = {M^2(2)}_{ll} + e \;\sin^2 \theta_w\;
m_z^2 \;\cos 2 \beta$ and $A=A_{0ll} - \mu_0 \tan \beta$ 
with $T_3 = -1/2$ and $e=-1$ for the down type
charged sleptons. ${M^2(1)}_{ll}, {M^2(2)}_{ll}$ and $A_{0ll}$  
are $R$ parity
conserving soft supersymmetry breaking parameters. The charged
slepton mass eigenstates are given by
\bea
{\tilde l}_i = U^{\tilde l}_{i1} {\tilde l}_L + U^{\tilde l}_{i2}
{\tilde l}_R  
\eea
The mixing angle $\phi = \phi_{l_k}$ (where $k$ correspond to the
particular generation we are considering) for the charged slepton is given by
\bea
\sin 2 \phi_{l_k} = { 2 A m_l  \over \sqrt{{(L^2 -R^2 )}^2+ 4 A^2 m_l^2} }
\eea
For $L-R$ mixing of $d$-squark we replace the the slepton mixing
angle $\phi_{l_k}$ by $\phi_{d_k}$ which is obtained similarly from the
above equations after replacing  $m_l$, $e=-1$, 
${M^2(1)}_{ll}, {M^2(2)}_{ll}$ and
$A_{0ll}$ with $m_d$, $e=-1/3$, 
${M^2(1)}_{dd}, {M^2(2)}_{dd}$ and
$A_{0dd}$ respectively. Similarly for $L-R$ mixing of $u$-squark
we replace the the slepton mixing
angle $\phi_{l_k}$ by $\phi_{u_k}$ which is obtained similarly from the
above equations after replacing  $m_l$, $T_3=-1/2$, $e=-1$, 
${M^2(1)}_{ll}, {M^2(2)}_{ll}$ 
with $m_u$, $T_3 = 1/2$, $e=2/3$, 
${M^2(1)}_{uu}, {M^2(2)}_{uu}$ respectively. Furthermore, in this
case $A = A_{0uu} -\mu_0 \cot \beta$.
For large $\mu_{\a}$ one should also include a term $-\mu_i
\l_{ijk} v_u /(\sqrt{2} m_l)$ for slepton and a term $-\mu_i
\l^{\pr}_{ijk} v_u /(\sqrt{2} m_d)$ for $d$ squark \cite{kong}.

Due to trilinear $\l_{ijk}$  couplings 
AMMM gets conribution from one loop diagram with (a) charged lepton
and sneutrino (b) charged slepton and neutrino, in the internal
lines. Photon line is attached with charged internal particle
line. Due to $\l^{\pr}_{ijk}$ couplings AMMM gets contribution
from one loop diagram with (a) up type quark and down type
squarks (b) down type quark and up type squarks, in the internal
line. Photon line is attached with any one of the internal
particle line. The chirality flip on the
external muon line or the chirality flip in the internal
sfermion line can be considered for both $\l$ and $\l^{\pr}$ 
couplings. However, while considering the
chirality flip on the internal line  one requires $L-R$ mixing of
sfermions. Then it will be difficult to get appreciable $\Delta
a_{\mu}$ as well as small neutrino mass $\sim 2.2$ eV as $L-R$
mixing of sfermions is present in both $\Delta a_{\mu}$ and
neutrino mass.
This makes the contribution to $\Delta a_{\mu}$ from the
chirality flip on the internal line lesser
than that coming from the chirality flip on the external muon
line and the contribution to
$\Delta a_{\mu}$ coming from the chirality flip on the internal
sfermion line may be ignored. 
If we consider the chirality flip on the
external muon line \cite{kim,leve} then  
for
$\l_{ijk}$ couplings 
\bea
\Delta a_{\mu}  \approx \sum_{i,k} {m_{\mu}^2  \over 96 \pi^2}
\left[  2 \;{{\mid \l_{i2k}
\mid}^2 \over m_{\tilde{\n_i}}^2 } + 2 \;{{\mid \l_{ik2}
\mid}^2 \over m_{\tilde{\n_i}}^2}  - \;{{\mid \l_{ik2}
\mid}^2 \over m_{\tilde{l_i}}^2} - \;{{\mid \l_{i2k}
\mid}^2 \over m_{\tilde{l_i}}^2} \right]      
\eea
and 
due to $\l^{\pr}_{ijk}$ couplings 
\bea
\Delta a_{\mu}  & \approx & \sum_{i,j} { m_{\mu}^2 \;{\mid \l^{\pr}_{2jk}
\mid}^2 \over 32 \pi^2
\left( m_{\tilde{d_{kR}}}^2-  m_{u_j}^2 \right) } \left[ 1+ {2 \;
r(u_j,d_{kR}) \over   1 - r(u_j,d_{kR})  }  \left\{ {1 \over 2} +
 { 3 \over 1 - r(u_j,d_{kR})} \right. \right. \nonumber \\ 
& + & \left. \left.  {2 + r(u_j,d_{kR}) \over 1 -
r(u_j,d_{kR}) } \;\ln r(u_j,d_{kR})  \right\} \right]
\eea
where $r(u_j,d_{kR}) = {\left(m_{u_j}/m_{d_{kR}} \right)}^2$. 
Particularly for $j=3$
and $k=1$ AMMM gets contribution through $\l^{\pr}_{231}$
coupling. For 
$m_{\tilde{d_3}} \sim 1   $ TeV one may consider
$\l^{\pr}_{231} \sim 1 $. For  $m_{\tilde{d_1}} \sim 200   $ GeV
one gets  $\Delta a_{\mu} \sim 10^{-9}$. Due to the stringent
constraint on other $\l^{\pr}_{2jk}$ couplings \cite{barg} it is
difficult to get significant contribution to $\Delta a_{\mu}$. As
for example  $\l^{\pr}_{231}$ coupling (which has been considered
in ref. \cite{kim}) may be considered about 1
but in that case $b$ squark has to be considered in the TeV range
to satisfy the present experimental bound \cite{barg} but as its' mass also
appear in the denominator it is not possible to get appreciable
contribution to $\Delta a_{\mu}$. For $\l_{i2k}$ or $\l_{ik2}$
couplings  it is possible to get appreciable contribution to
$\Delta a_{\mu}$ for various values of $i$ and $k$. If one
considers the mass of charged slepton in the TeV range then most
of these couplings can be about 1 satisfying the present
experimental constraint \cite{barg}. In this case if one
considers the mass of sneutrino somewhat light (about 100 GeV
say) then $\Delta a_{\mu} \sim 10^{-9}$ and also it is
interesting to note that the negative contributions are very
small due to high charged slepton mass. 

We like to mention here  why the diagrams contributing to AMMM
with chirality flip on the internal line are small. With
chirality flip in $u$ squark or $d$ squark internal particle line
the one loop diagram gives  
\bea
\Delta a_{\mu} &=&  \sum_{j,k} { N_c  \;m_{\mu}  \over 8 \pi^2 
m_{\tilde{ d}}^2 }
\;\l^{\pr}_{2jk} \;\l^{\pr}_{2kj}  \left[ m_{d_j} \;\sin 2 \phi_{u_k}  \left\{
{2 \over 3} F_1(r_{d_j})  + {1 \over 3} F_2(r_{d_j})  \right\}
\right. \nonumber  \\
& + & \left.  
m_{u_j} \;\sin 2 \phi_{d_k}  \left\{ { 1 \over 3 } F_1(r_{u_j})  + 
{ 2 \over 3 } F_2(r_{u_j})  \right\} \right]
\eea 
where 
\bea
F_1(r_{d_j,u_j})  = {1 \over 2 {(1-r)}^2} \left[ 1 + r + {2 \;r \;\ln r
\over 1 - r}\right]
;\;\;\;
F_2(r_{u_j})  = {1 \over 2 {(1-r)}^2} \left[ 3 - r + {2 \;r \;\ln r
\over 1 - r}\right]
\nonumber \\
\eea
and $r_{d_j,u_j} = {\left(m_{u_j,d_j}/m_{\tilde q} \right)}^2$
and $m_{\tilde q}$ is the scalar squark mass in the loop.
However, after diagonalising the $L-R$ mixing matrix there are
two squarks $\tilde {q_1}$ and $\tilde {q_2}$ in the internal line. 
So there will be further
suppression  in the above $\Delta a_{\mu}$ by a factor $\left( m_{\tilde
{q_2}}^2 - m_{\tilde
{q_1}}^2 \right) / \left( m_{\tilde
{q_1}}^2 m_{\tilde
{q_2}}^2 \right)$. 
It is possible to get appreciable $\Delta a_{\mu}$ due to top mass 
which comes for $j=3$.  However, in that
case  $\Delta a_{\mu}$ is also proportional to $L-R$ mixing of
$d$ squark. But this mixing will also be present in the neutrino
mass matrix. So it will not be possible to get small neutrino
mass as well as large $\Delta a_{\mu}$. For $\l_{ijk}$ couplings
with chirality flip on the internal line there
will be  diagram contributing to $\Delta a_{\mu}$ with charged
slepton and neutrino in the internal line. However, as $\Delta
a_{\mu}$ as well as neutrino mass - both are proportional to 
$L-R$ mixing of charged lepton, such diagrams cannot give
appreciable $\Delta a_{\mu}$. So we conclude that the
contributions to $\Delta a_{\mu}$ coming from the chirality flip
on the internal line can be ignored.

We next like to show what happens to neutrino mass \cite{hall} if one
considers such higher values of $\l_{ijk}$ or
$\l^{\pr}_{ijk}$ couplings of about 1. 
As we have ignored the mixing of neutrinos with neutralinos,
neutrino does not get mass at the tree level. However, at one
loop level all the neutrinos will acquire significant 
mass as the trilinear
couplings are  required to be large for AMMM.
From Figure 1, one loop contribution to the neutrino mass matrix 
${(m_{\nu})}_{ij}$ due to $\l^{\pr}_{ijk}$ couplings
is given by  
\bea
{(m_{\nu})}_{ij}= {3 \over 32 \pi^2} \sum_{n,k} \l^{\pr}_{ikn} 
\;\l^{\pr}_{jnk} \;m_{d_k} \;\sin 2 \phi_{d_n} \;\ln{m_1 \over m_2}
\eea
where $m_{1,2}$ are the non-degenerate squark masses obtained
after diagonalising the the squark mass matrix inducing $L-R$ mixing.
Similarly, for $\l_{ijk}$ couplings the one loop contribution to
the neutrino mass matrix is given by
\bea
{(m_{\nu})}_{ij}= {1 \over 32 \pi^2} \sum_{n,k} \l_{ikn} 
\;\l_{jnk} \;m_{l_k} \;\sin 2 \phi_{l_n} \;\ln{m_1 \over m_2}
\eea
For $\l_{ijk} \sim 1 $ or  $ \l^{\pr}_{ijk} \sim 1$ considering
the upper bound on $m_{\n}$ as 2.2 eV one obtains the following
bound on the $L-R$ mixing of sfermions: 
\bea
\sin 2 \phi \;\ln{m_1 \over m_2} \leq  6.3 \times 10^{-4}/m_{l_{j,k}} 
\eea
where $m_{l_{j,k}}$ is the mass of the charged lepton of
generation $j$ or $k$. Particularly for $j=3$ or $k=3$ for $\l_{ijk}$
or $\l^{\pr}_{ijk}$ couplings this bound is highly stringent
(about $3.7 \times 10^{-7}$). This means that the $A$ parameter should
be very small compared to $\sqrt{L^2-R^2}$ and that implies
particularly for large $\tan \b >> 1$ (which is theoretically
preferred particularly for bottom-tau unification) the $\mu_0$ parameter
should be very small. So certain amount of fine tuning is
necessary as this parameter is present in unbroken supersymmetry.
If one considers the $R$ parity violating contribution to $A$
parameter \cite{kong} as mentioned earlier then it indicates
other $\mu_{i}$ also should be small.  

Next we shall show that we do not 
require such significant suppression of $L-R$
mixing to get appreciable $\Delta a_{\mu}$ as
well as small neutrino mass if we consider horizontal symmetry
for the lepton doublet.  For this  we consider the $R$
parity violating Supersymmetric Model in such a way that   $l_e
-l_{\mu}$ lepton number  is unbroken and there is discrete
horizontal symmetry $D$ between  the first two generations of
lepton \cite{babu}.  Under    symmetry $D$  the following chiral
superfields   transform as
\bea
\left(
\begin{array}{c} L_e  \\
L_{\mu}
\end{array}
\right)  \rightarrow \left(
\begin{array}{cc} 0 & 1 \\
-1 & 0
\end{array}
\right)
\left(
\begin{array}{c} L_e  \\
L_{\mu}
\end{array}
\right). 
\eea

\bea
\left(
\begin{array}{c} e^c  \\
{\mu}^c
\end{array}
\right)  \rightarrow \left(
\begin{array}{cc} 0 & 1 \\
-1 & 0
\end{array}
\right)
\left(
\begin{array}{c} e^c  \\
{\mu}^c
\end{array}
\right). 
\eea
The superfields $L_{\tau}, \tau^c , Q_i, u_i^c$ and $d_i^c$ do
not transform under $D$. Under $l_e-l_{\mu}$ and $D$ 
the most general superpotential in Eqn.
(1) now takes the form
\bea
W&=&\e_{ij}  \left[ -\mu_{\a} L_{\a = 0,3}^i H_U^j + {1 \over 2}
\left\{ \l_{1
31} L_e^i L_{\tau}^j e^c  + \l_{2
32} L_{\mu}^i L_{\tau}^j \mu^c + \l_{1
23} L_e^i L_{\mu}^j \tau^c +  \l_{0
33} L_{\tau}^i H_d^j \tau^c \right. \right. \nonumber \\ 
&+& \left. \left. \l_{0
11} L_e^i H_d^j e^c + \l_{0
22} L_{\mu}^i H_d^j \mu^c  \right\}   + \l_{0 n m }^{\pr} H_d^i Q_n^j
D_m +\l_{3 n m }^{\pr} L_{\tau}^i Q_n^j
D_m  \right. \nonumber \\  &-& \left. h_{nm} H_U^i Q_n^j U_m \right]
\eea
{\it It is important to note here that there are only a few trilinear
$\l$ and $\l^{\pr}$ couplings and there will be no contribution
to AMMM from $\l^{\pr}$ coupling}. Now $l_{\tau}$  and $H_d$ have
identical transformation properties and apart from two Higgs vev
there is tau sneutrino vev $v_3$. One can choose the appropriate
basis so that $v_3 = 0$.  $v_1$ and $v_2$ do not exist  because
of $l_e-l_{\mu}$  conservation.
If $D$ symmetry is considered for the soft breaking terms then
instead of Eqn. (2) one can write down similar terms like above
for $V_{soft}$. Particularly $D$ symmetry will imply ${\left( 
M_{\tilde L}^2 \right)}_{11}  = {\left( 
M_{\tilde L}^2 \right)}_{22} $  and ${\left( M_{\tilde E}^2
\right)}_{11}  = {\left( M_{\tilde E}^2
\right)}_{22} $. After breaking of gauge symmetry $D$ symmetry
can be kept unbroken. 
Although in the $D$ symmetric limit the neutrino mass for the
1-st two generations are zero, the electron and muon mass will
also be
degenerate. However, as has been discussed in ref. \cite{babu},
by adding dimension two soft breaking terms 
for muon superfields breaking $D$ symmetry softly
it is possible to get appropriate mass differences for electron
and muon without affecting the neutrino mass spectrum. 
It is because instead of lepton mass entering the neutrino mass diagram
it is the gaugino mass (having much larger magnitude) 
that enters into the charged lepton mass 
diagram.

In the $D$ symmetric limit neutrino mass terms for the 1-st two
generations are forbidden. Only $\l_{131} = - \l_{311}$ and
$\l_{232} = - \l_{322}$ will give mass to $\tau$ neutrino. These
$\l$ couplings will contribute to AMMM , however those
contributions may be very small. On the other hand, $\l_{123}$
coupling will not contribute to neutrino mass but will contribute
to AMMM as shown in Figure 2. This will give
\bea
\Delta a_{\mu} =  { m_{\mu}^2 \;{ \mid \l_{123} \mid }^2 
 \over 48 \pi^2} \left[  
{1  \over m_{\tilde{\n_e}}^2} -  {1  \over 2 m_{\tilde{\tau}}^2} \right]
\eea
from the diagram with (a) $\tau$ lepton and $\tilde{\n_e}$ in the
internal line (shown in Figure 2) and (b) $\tau$ slepton and $\n_e$ in 
the internal line (not shown in figures). 
Photon is attached with $\tau$ lepton for (a) and $\tau$ slepton
for (b) and there is
chirality flip on the external muon line.   In $R$ - violating
Supersymmetry the lower bound on sneutrino mass is expected to be
somewhat lower than 100 GeV due to sneutrino pair production and
its subsequent decays to charged leptons at LEP \cite{abreu}. 
For  $m_{\tilde{\tau}}$ 
about a few   TeV  it is possible
to consider $\l_{123} \sim 1$ \cite{barg}.
In that case 
the negative contribution to AMMM is very small
and considering the conservative bound $m_{\tilde{\n_e}}
\sim 100 $ GeV  one gets $\Delta a_{\mu} \sim 10^{-9}$.

In this scenario due to $l_e-l_{\mu}$ symmetry the rare processes
like $\mu  \rightarrow 3 e$,  $ \mu \rightarrow e \gamma $, $\tau
\rightarrow 3 e $, $ \tau \rightarrow \mu ee$ etc. are forbidden.
However,    due to $\l_{123}$ coupling  $\mu$ will decay to $e$,
$\bar{\n_e} $, $\n_{\mu}$ through $\tau$-slepton exchange diagram
at the tree level apart from the Standard Model $W$ exchange
diagram.   So there will be $e-\mu-\tau$ universality violation.
Our studies indicate that this violation might be 
observed in near future due to our requirement of higher value of
$\l_{123}$ coupling  (for $\tau$-slepton mass about a few TeV ) to 
explain the presently observed $\Delta
a_{\mu}$ with $l_e-l_{\mu}$ and $D$ symmetry  in $R$ parity
violating Supersymmetric Model. 

There are some uncertainties in the calculation of the hadronic
contribution to AMMM \cite{marci,uncer} particularly in the dispersion
integral approach to hadronic vacuum polarization effects 
and as such the uncertainties
also exist in finding the amount of what should be the new physics
contribution to AMMM. To satisfy the experimental data
\cite{brown} the hadronic and the new physics contribution
together \cite{marci} to AMMM should be $7350 (153) \;\;10^{-11}$. Total
hadronic contribution as shown by Davier {\it et al} is $6294(62) \; 
10^{-11}$ whereas the same as shown by Jegerlehner is $6974(105)
\; 10^{-11}$. If the later is correct then it is easily possible
to explain $\Delta a_{\mu}$ through $R$ parity violating
interactions. However, if the hadronic contribution is really
less then one have to carefully examine the possiblity of the
explaination of  $\Delta a_{\mu}$ 
in terms of R parity violating interactions.

Lastly we like to comment on the AMMM from $R$ parity conserving
Supersymmetric model versus $R$ parity violating Supersymmetric
Model. 
In the $R$ parity conserving case
there will be diagrams a) with smuon and neutralino b)
sneutrino and chargino, in the internal lines. 
If all the supersymmetric particles
present in the loop are
considered almost degenerate then the diagram with (b) will
dominate otherwise for large $L-R$ mixing of smuons in some cases
diagram with (a) may dominate depending on the right-handed smuon
mass. However, in both the cases large
$\tan \b$ is preferred as that gives larger contribution to AMMM.
For theoretically preferred value of $\tan \b \sim 35$
(necessary for the bottom-tau unification) the
maximum mass allowed is around 360 Gev for the chargino and/or
neutralino and 420 Gev for the lightest slepton in the loop \cite{msusy}.
However, in the $R$-parity violating case  as there is lepton
number violation, Majorana neutrinos can be massive. To satisfy
the upper bound on neutrino masses it seems natural to consider
$l_e-l_{\mu}$ and $D$ symmetry. Then 
we require light $\tilde{ \n_e}$ mass $\sim 100 $GeV
but other slepton masses are rather heavy in the TeV range. Here,
it is interesting to note 
that theoretically sneutrino could be the lightest
supersymmetric particle. 

\newpage
\hspace*{\fill}

{\bf Acknowledgment}

The authors would like to thank E. Ma for his comments and for
reading the manuscript and would like to thank 
B. Mukhopadhyaya for providing useful 
information. 
One of us (RA) is supported by D.S.T. India. RA thanks the
Institute of Mathematical Sciences for hospitality at the start
of this study.

\newpage
\begin{figure}[htb]
\mbox{}
\vskip 1.2in\relax\noindent\hskip 0.8in\relax
\includegraphics{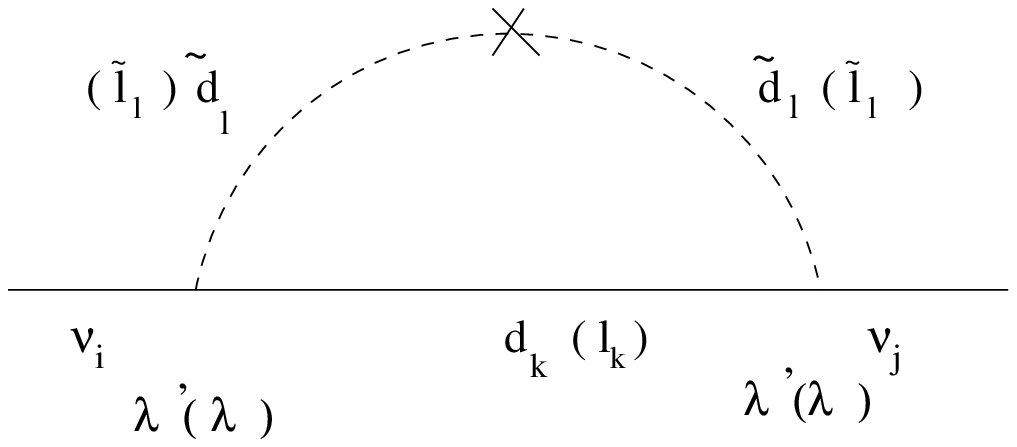} \vskip .25in
\caption{   One loop  diagram  involving $L$-violating  couplings  
  generating neutrino 
mass.}
\end{figure}
\begin{figure}[htb]
\mbox{}
\vskip 2.4in\relax\noindent\hskip 0.8in\relax
\includegraphics{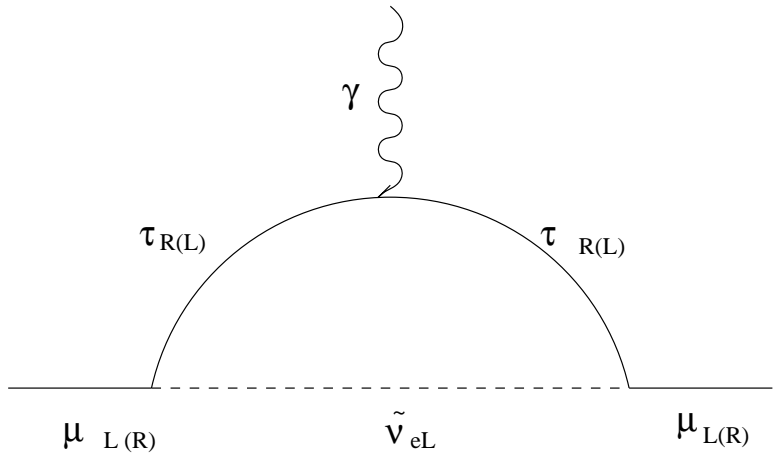} \vskip .25in
\caption{   One loop  diagram  involving $L$-violating  couplings  
  generating anomalous magnetic moment of muon 
mass.}
\end{figure}

\end{document}